\documentclass[showpacs,preprint,aps,floats,superscriptaddress]{revtex4}
\usepackage{amsmath}
\usepackage{graphicx}
\usepackage{epsfig}
\usepackage{color}
\usepackage[latin1]{inputenc}

\begin{document}
\newcommand{\dg}{^{\dagger }}
\newlength{\upit}\upit=0.1truein
\newcommand{\raiser}[1]{\raisebox{\upit}{#1}}
\newlength{\bxwidth}\bxwidth=1.5 truein
\newcommand\frm[1]{\epsfig{file=#1,width=\bxwidth}}
\def\fig#1#2{\includegraphics[height=#1]{#2}}
\def\figx#1#2{\includegraphics[width=#1]{#2}}
\newlength{\figwidth}
\newcommand{\fg}[3]
{
\begin{figure}[ht]
\[
\includegraphics[width=\figwidth]{#1}
\]
\vspace*{-4mm}
\caption{\label{#2}
\small
#3
}
\end{figure}}
\newcommand{\mat}[4]{\left[
\begin{array}{cc}
#1 & #2 \cr #3 & #4
\end{array} \right]
}
\newcommand{\cmat}[4]{\left (
\begin{array}{cc}
   #1 & #2 \cr #3 & #4
\end{array} \right)
}
\newcommand{\beq}{\begin{equation}}
\newcommand{\eq}{\end{equation}}
\newcommand{\bea}{\begin{eqnarray}}
\newcommand{\ea}{\end{eqnarray}}
\newcommand{\nn}{\nonumber}

\title
{\bf Entanglement generation in a system of two atomic
quantum dots coupled to a pool of interacting bosons}

\author{Anna Posazhennikova}
\email[Email: ]{anna.posazhennikova@rhul.ac.uk}

\affiliation{Department of Physics, Royal Holloway, University of London,
  Egham, Surrey TW20 0EX, United Kingdom}
\affiliation{Fachbereich Physik, Universit\"at Konstanz, D-78457, Konstanz, Germany}

\author{Reinhard Birmuske}
\affiliation{Fachbereich Physik, Universit\"at Konstanz, D-78457, Konstanz,
Germany}

\author{Martin~Bruderer}
\affiliation{Institut f\"ur Theoretische Physik, Albert-Einstein Allee 11,
Universit\"at Ulm, 89069 Ulm, Germany}

\author{Wolfgang Belzig}
\affiliation{Fachbereich Physik, Universit\"at Konstanz, D-78457, Konstanz, Germany}

\preprint{version of \today}


\begin{abstract}
We discuss entanglement generation in a closed system of one
or two atomic quantum dots (qubits) coupled via Raman transitions
to a pool of cold interacting bosons. The system exhibits rich entanglement
dynamics, which we analyze in detail in an exact quantum mechanical
treatment of the problem. The bipartite setup of only one atomic
quantum dot coupled to a pool of bosons turns out to be equivalent to two
qubits which easily get entangled being initially in a product state. We
show that both the number of bosons in the pool and the
boson-boson interaction crucially affect the entanglement
characteristics of the system. The tripartite system of two atomic
quantum dots and a pool of bosons reduces to a qubit-qutrit-qubit
realization. We consider entanglement possibilities of the pure system as
well as of reduced ones by tracing out one of the constituents, and show how
the entanglement can be controlled by varying system parameters. We
demonstrate that the qutrit, as expected, plays a leading role in
entangling of the two  qubits and the maximum entanglement depends in a
nontrivial way on the pool characteristics.

\end{abstract}


\maketitle


\section{Introduction}

Entanglement is a hallmark of quantum mechanics and is fundamentally
different from any correlation known in classical physics. It has become
indispensable in quantum computation because of its enormous capabilities to
process information in novel ways \cite{Nielsen}. Entangled states form a
base for quantum communication protocols such as 
superdense coding \cite{superdence} or quantum teleportation
\cite{teleport}.

Entanglement also plays a fundamental role in the physics of condensed
matter and strongly correlated materials \cite{Luigi}. BCS wave function
\cite{BCS}, Laughlin ansatz \cite{Laughlin}, Kondo singlet \cite{Hewson} are
examples of highly entangled ground states. Entanglement entropy provides
significant insight into quantum critical phenomena
\cite{Osterloh,Vidal,Refael} and is used to characterize topological order
\cite{Kitaev,Wen}, which is not described by the standard Ginzburg-Landau
theory.

It would be interesting for both quantum information and condensed matter
if one could generate particular entangled states in a controlled manner.
Cold atoms in optical lattices have become one of the favorite systems to
tailor various many body phenomena, including multi-particle entanglement
\cite{Mandel}, which may be used for quantum computing
\cite{Briegel}. Schemes based on arrays of cold atoms have also been
proposed for generating highly entangled cluster states and entangling gates
between distant qubits~\cite{Clark1,Clark2,Bose}. The main advantage of cold
atomic systems is that they are clean, well
controlled and almost dissipationless. Moreover, impressive 
recent advances in fluorescence imaging have made it possible to probe with
high fidelity the on-site number statistics of cold atoms in optical
lattices in the Mott regime \cite{Bakr,Sherson}.

In this paper we consider the emergence of entanglement in a system of two
atomic quantum dots (AQDs), which constitute two qubits, coupled
by optical transitions to a pool of bosons containing a finite number
of particles. The bosons in the pool are interacting, and we consider
the ratio of their interaction energy $U$ with respect to the
coupling $T$ between the AQDs and the bosons as a main parameter in our
problem. We first consider an even simpler system of just one AQD
coupled to the pool and show that such a set-up is in fact
equivalent to a two-qubit system whose ground state and
excited state are a singlet and triplet state, respectively. We observe
dynamic formation of entanglement in the system, starting initially from a
product state. We show that the state can become maximally entangled in
the regime $U/T\ll1$.

We then proceed to the case of two AQDs coupled to the pool. Since
the system is closed and the number of particles is conserved, this
is equivalent to a qubit-qutrit-qubit realization, which is interesting
because qutrits have larger entanglement capacitance than qubits
\cite{Wootters2001}. Although there is no well-defined entanglement measure
for a tripartite system, we calculate a concurrence which tells us
whether or not the state is entangled. We then trace out the
qutrit (pool of bosons) and calculate entanglement of
formation~\cite{Wootters98} of the bipartite system. We demonstrate
that dynamics of the entanglement of formation is consistent with that of
the concurrence and discuss best entangling possibilities depending on
$U/T$. Finally, we trace out one of the qubits and calculate both
entanglement of formation and negativity \cite{Werner02} for a remaining
qubit-qutrit state. 

Our work is further motivated by a previous result by some of us \cite{EPL}, in which we found 
that strong correlations can emerge between the two qubits coupled to a Bose-Einstein condensate. 
However, we have not found any entanglement in that case, most likely since the Bose-Einstein condensate 
was described by the Gross-Pitaevskii wave-function, which is effectively a classical field. 
The question we want to address here is therefore, whether quantum entanglement emerges
 if the BEC is not in the Gross-Pitaevski regime.


\section{Model: Two AQDs coupled to a pool of interacting bosons}

Our set-up is displayed in Fig. 1. We consider a system of two atomic
quantum dots coupled to a pool of bosons trapped in an external potential
$V_{bos}$.  The system Hamiltonian reads \beq H=H_{bos}+H_{AQD}+H_{mix}, \eq
The energy levels of $V_{bos}$ are widely separated, so that particles can
only occupy the lowest-lying level. This is effectively a one-site
optical lattice in the Mott regime, so that $H_{bos}$ reads \beq H_{bos}=E_b
n+\frac{U}{2}n(n-1), \label{bosons} \eq where $E_b$ is the energy of
the bosons, $n$ is the number operator and $U$ is the interaction
energy between the atoms.

Each atomic quantum dot constitutes a qubit and can be created in the
following way: An atom of a different hyperfine species from
the bosons in $V_{bos}$ is trapped in a tight potential $V_{1,2}$. The
interaction between these atoms is assumed to be so large that double
occupancy of the lowest lying level of $V_{1,2}$ is forbidden. The coupling
between qubit and bosons in the pool is produced in an optical way suggested
in \cite{Recati}: An external blue-detuned  laser induces Raman
transitions between the dot and the atoms in the pool. The possibility of
such a scheme was also confirmed in a numeric simulation \cite{Morigi}.
 Importantly, the magnitude of the coupling between different
atomic states can be rather large, since it is controlled by adjustable Rabi
frequency of the external laser. However, the overlap between the
wave-functions of the bosons in the trap $V_{bos}$ and bosons of atomic
quantum dots is considered to be small, so that density-density interaction
between particles can be neglected.

Under above-mentioned conditions we can map the bosonic creation
(annihilation) operators $b^\dag$ ($b$) of the dot onto their
Pauli matrix equivalents: $b^{\dg}\rightarrow \sigma_{+}$ and $b\rightarrow
\sigma_{-}$. The corresponding Hamiltonian is then
\beq H_{AQD}=-\frac{\Delta}{2}\sum_{i=1}^{2}(1+\sigma_z(i)), \eq
The argument of the Pauli matrix $i$ is referring to the left $i=1$ or
the right $i=2$ qubit. Here, $\Delta$ is the detuning energy necessary
to avoid spontaneous emissions during the Raman process.

\begin{figure}[!bt]
\begin{center}
\hspace*{1em}
\includegraphics[width=0.45\textwidth]{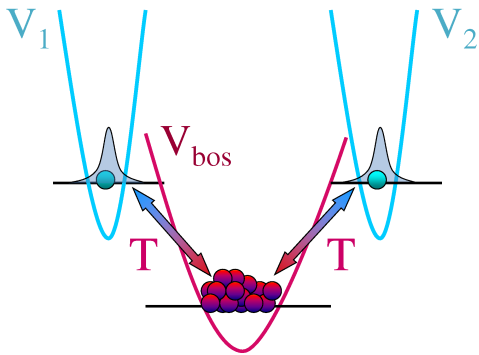}\vspace*{-0.5em}
\end{center}
\caption{
Two atomic qubits coupled to a finite number of cold interacting bosons. The
qubits are trapped by the tight potentials $V_1$ and $V_2$, the bosons are
confined in an external potential $V_{bos}$. $T$ is the optical transition
assisted coupling between the qubits and the bosons.}
\label{setup}
\end{figure}

Finally, the interaction between the qubits and interacting bosons in the
pool reads
\beq
H_{mix}=T\left[a\sigma_{+}(1)+a\dg\sigma_-(1)+a\sigma_{+}(2)+a\dg\sigma_-(2)
\right], \eq
where $a^\dag$ ($a$) creates (annihilates) a boson in the pool and the
coupling $T$ is proportional to the Rabi frequency $\Omega_R$ of the
above-mentioned Raman transition, i.e., $T \sim \hbar\Omega_R$. We
consider a closed system and the total particle number
$N=n+\frac{1}{2}(1+\sigma_z(1))+\frac{1}{2}(1+\sigma_z(2))$ is
conserved, i.e., $[H,N]=0$. In the following we tackle the problem of
the dynamics of entanglement in this tripartite system.


\section{A single AQD coupled to the pool: Bipartite entanglement}

It is instructive to start from an even simpler bipartite case of only one
AQD coupled to the pool of bosons. Due to the particle number
conservation the corresponding Hilbert space
$\mathcal{H}=\mathcal{H}_{bos}\otimes \mathcal{H}_{AQD}$ is spanned just by
two orthonormal states
\bea
|n,0\rangle&=&|n\rangle \otimes |0\rangle , \nn \\
|n-1,1\rangle&=&|n-1\rangle \otimes |1\rangle. \label{basis_1} \ea
The state $|0\rangle$ corresponds to ``spin-down'' state of a dot, while the
state $|1\rangle$ corresponds to a ``spin-up''. The wave function in this
case is \beq |\Psi(t)\rangle =c_1(t) |n,0\rangle + c_2(t)|n-1,1\rangle
\label{wf_two} \eq with the normalized coefficients
$|c_1(t)|^2+|c_2(t)|^2=1$. The wave function \eqref{wf_two} is in fact
equivalent to a standard wave function of two spins-1/2 or two qubits, with
two of the four coefficients equal to zero from the onset. We can
immediately write down the time-dependent concurrence $C(t)$ of
the state as $C(t)=2|c_1(t)c_2(t)|$ and use it in the following as
an entanglement measure \cite{Wootters98}. The state is entangled
whenever both coefficients $c_1$ and $c_2$ are nonzero. Note that
the initial state $|\Psi(0)\rangle= |n,0\rangle$ is always
unentangled.

The Hamiltonian in the basis \eqref{basis_1} reads
\beq H_1=H'_1 \hat I_2 +\left (
\begin{array}{cc}
0 & T\sqrt{n} \cr
T\sqrt{n} & \quad -E -U(n-1)
\end{array} \right),
\label{ham_two} \eq
where $H'_1=E_b n +\frac{1}{2}Un(n-1)$ is the diagonal part of the
Hamiltonian (which we neglect in the further consideration) and
$E\equiv E_b+\Delta$ is  the energy
difference between the traps. The Hamiltonian \eqref{ham_two}
describes a two-level system with Rabi frequency $\omega_R=T\sqrt{n}$ and
detuning $\delta=\frac{1}{2}[-E-U(n-1)]$, so that the
eigenenergies are $E_{\pm}=\delta\pm\sqrt{\delta^2+\omega_R^2}$. We
note that for $\delta=0$ the ground state is a singlet and the excited
state is a triplet.

We now consider the time evolution of the concurrence $C(t)$,
or equivalently, the entanglement of the state \eqref{wf_two},
initially unentangled $|\Psi(0)\rangle= |n,0\rangle$. First, we discuss
the case of small detuning $\delta/T\ll 1$. The maximum
concurrence is then always close to unity and the concurrence is
approximately described by $C\sim |\exp(-2i\delta t)
\sin(2\sqrt{n}t)|$, shown in Fig. 2({\it a}) for various $U/T$ and fixed
$E$ and $n$. In order to analyze the results, we introduce a convenient
quantity $t_{ent}$ referred to as "entanglement period", which is the
time span between two subsequent instances of zero entanglement. In
Fig. 2({\it a}), for example, $t_{ent}=0.5$ for $U/T=0$ and
$t_{ent}=1$ for any finite $U/T$ (time is in the units of $1/T$).

In Fig. 2 ({\it b}) we plot the entanglement
period and the maximum concurrence $C_{max}$ as a function of the
interaction $U/T$ for two values of $n$. We see from the results in Fig. 2 ({\it
b}) in order to increase the entanglement period and maximally
possible entanglement one should either decrease $U/T$ or the number of
particles in the bosonic pool $n$. Large values of $n$ as well as large
values of $U/T$ do not generally advocate entanglement in the system. This
happens because the amplitude of the Rabi
oscillations scales as $\omega_R/(\omega_R^2+\delta^2)$, so that large
values of $\delta$ contribute to the suppression of entanglement. 

\begin{figure}[!bt]
\begin{center}
\includegraphics[width=0.9\textwidth]{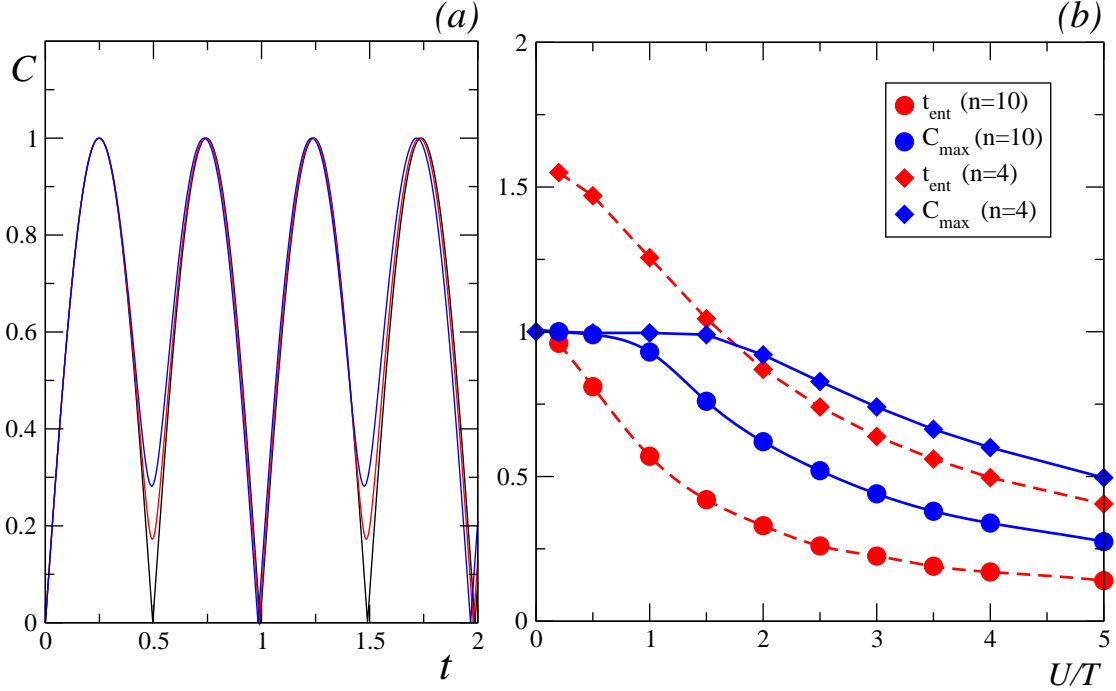}\vspace*{-0.5em}
\end{center}
\caption{
({\it a}) Time evolution of the concurrence of the bipartite setup for $n=10$, $E/T=0.01$ and the interaction
$U/T=0$ (black), $U/T=0.06$ (red) and $U/T=0.1$ (blue).
({\it b}) Dependence of the entanglement period $t_{ent}$ (red curves) and maximum concurrence $C_{max}$ (blue curves) on the
 interaction
$U/T$  for n=10 (circles) and n=4 (diamonds) bosons in the pool.
}
\end{figure}


\section{Two AQDs coupled to the pool: Bipartite and tripartite entanglement}

We consider now our initial tripartite system: two AQDs coupled to the
pool of interacting bosons. The corresponding Hilbert space is a
tensor product of three subspaces $\mathcal{H}=\mathcal{H}_{bos}\otimes
\mathcal{H}_{AQD_1}\otimes \mathcal{H}_{AQD_2}$ and is spanned only by
four basis vectors (because of particle conservation) which we choose in the
following way
\beq \{|n+1,00\rangle, |n,01\rangle, |n,10 \rangle, |n-1,
11\rangle \}. \label{basis} \eq
Note that the pool is effectively a three state system, i.e., a
qutrit, described by the states ``spin-up'' $|n+1\rangle$, ``spin-flat''
$|n\rangle$ and ``spin-down'' $|n-1\rangle$.

We can now write the Hamiltonian of the tripartite system in
the basis \eqref{basis} as
\beq H=H_0 \hat I_4 +\left (
\begin{array}{cccc}
E+Un & T\sqrt{n+1} & T\sqrt{n+1} & 0 \cr
T\sqrt{n+1} & 0 & 0 & T\sqrt{n} \cr
T\sqrt{n+1} & 0 & 0 & T\sqrt{n} \cr
0 & T \sqrt{n} & T\sqrt{n} & -E+U(1-n)
\end{array} \right),
\label{ham} \eq
where $H_0=E_b n+\frac{1}{2}Un(n-1)-\Delta$ is the diagonal part. One
of the eigenstates of this Hamiltonian is a singlet in the AQD sector
$|\Psi_s\rangle = \frac{1}{\sqrt{2}} |n\rangle \otimes \left(|01\rangle
-|10\rangle \right)$ corresponding to an eigenvalue equal to zero. All
other eigenvalues have to be determined numerically from the matrix
\eqref{ham}. We just mention that the singlet
is the second excited state, which is very close to the third
excited state (the closer the smaller the interaction energy $U$ is).
The ground state and the third excited state are widely separated, the interlevel spacing being proportional to $2\sqrt{n}$ for small
$E/T$ and $U/T$. 


\subsection{Tripartite entanglement between the AQDs and the pool}

We will now turn to the issue of entanglement generation in the tripartite
system. The tripartite system is described by the pure state
\beq |\Psi\rangle=c_1|n+1,00\rangle+c_2 |n,01\rangle+c_3 |n,10 \rangle+c_4
|n-1, 11\rangle \label{wavefunc} \eq
where $c_i(t)$ are complex coefficient satisfying $\sum_{i=1}^4
|c_i(t)|^2=1$. Although an entanglement measure is not
unambiguously  defined for a tripartite hybrid system (one
qutrit and two qubits) one can calculate a concurrence \cite{Fei}
\beq C(|\Psi \rangle)=\sqrt{3-\sum_{i=1}^3 \mbox{Tr} \rho_i^2}
\label{concur_tri}. \eq
where $\rho_i$ is the density matrix of a subsystem $i$ with the two
other subsystems traced out. The system is unentangled if
this concurrence is zero, and a maximum of the concurrence
 corresponds to the entanglement maximum. We obtain from our full density matrix
$\rho=|\Psi\rangle \langle \Psi|$ the three reduced density matrices
\bea \rho_1&=&|c_1|^2|n+1\rangle \langle n+1|+(|c_2|^2+|c_3|^2)|n\rangle
\langle n|+
|c_4|^2 |n-1\rangle \langle n-1|, \\
\rho_2&=&(|c_1|^2+|c_2|^2)|0\rangle \langle 0|+(|c_3|^2+|c_4|^2)|1\rangle
\langle 1|,\\
\rho_3&=&(|c_1|^2+|c_3|^2)|0\rangle \langle 0|+(|c_2|^2+|c_4|^2)|1\rangle
\langle 1|,\ea
where $\rho_1$, $\rho_{2}$ and $\rho_{3}$ are the reduced density
matrices of the pool of bosons, the left AQD and the right AQD,
respectively. The resulting concurrence \eqref{concur_tri} reads
\beq C(|\Psi
\rangle)=\sqrt{3-3\sum_{i=1}^4|c_i|^4-2\left(|c_2|^2|c_3|^2+|c_2|^2|c_4|^2+|c_1|^2|c_2|^2+|c_1|^2|c_3|^2+|c_3|^2|c_4|^2\right)}\:.
\eq

\begin{figure}[!bt]
\begin{center}
\includegraphics[width=0.9\textwidth]{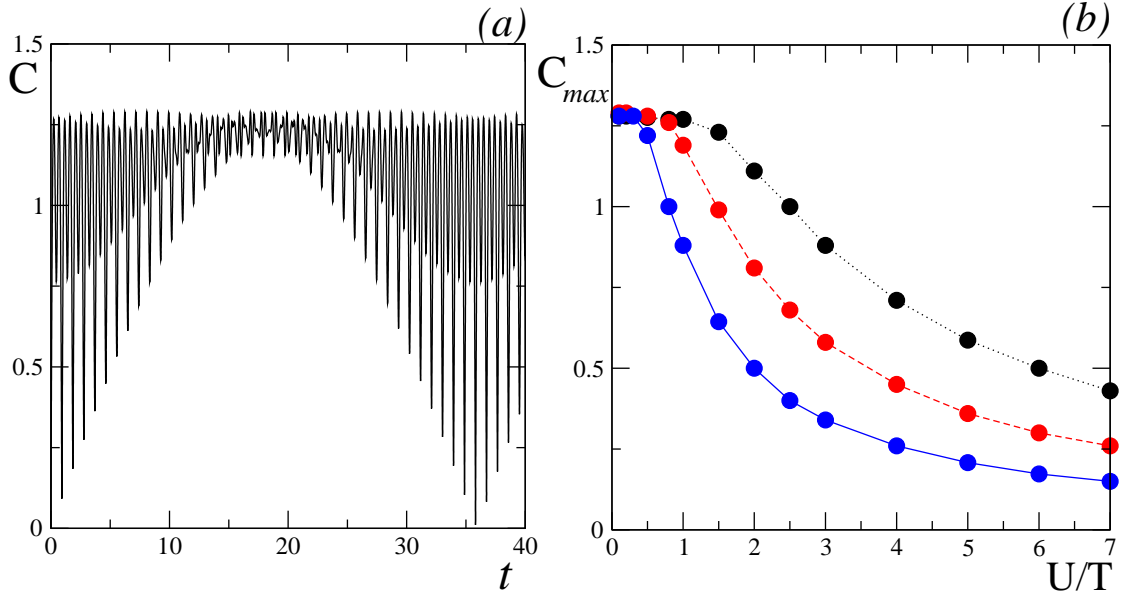}\vspace*{-0.5em}
\end{center}
\caption{\label{con_tri}
({\it a}) Time-dependent concurrence $C$ of the tripartite setup for $E/T=0.01$, $n=10$ and interaction $U/T=0.2$.
({\it b}) The maximum value of
concurrence $C_{max}$ of the tripartite system versus interaction $U/T$ for $n=4$ (black), $n=10$ (red) and
$n=30$ (blue) bosons in the pool. 
}
\end{figure}

In Fig. 3 ({\it a}) we display an example of the time-dependent
concurrence $C$ for $U/T=0.2$. The initial condition for Fig. 3
and the rest of the plots in the paper is $c_1(0)=1$, which corresponds
to two initially empty quantum dots. As expected, the time
dependence of the concurrence is more complicated than for the
bipartite case. However, it is again possible to identify an
entanglement period and the maximum concurrence over that period.
Note that the maximum concurrence is larger than $1$, which is not
surprising since our system comprises a qutrit---a quantum particle
with a larger entanglement capacitance than a qubit.  In Fig. 3 ({\it
b}) we plot the maximum concurrence versus interaction $U/T$ for various
$n$. We observe similar tendencies as in the previous bipartite case. The
maximum concurrence is suppressed by increasing either the interaction
or the number of bosons in the pool.

\begin{figure}[!bt]
\begin{center}
\includegraphics[width=0.8\textwidth]{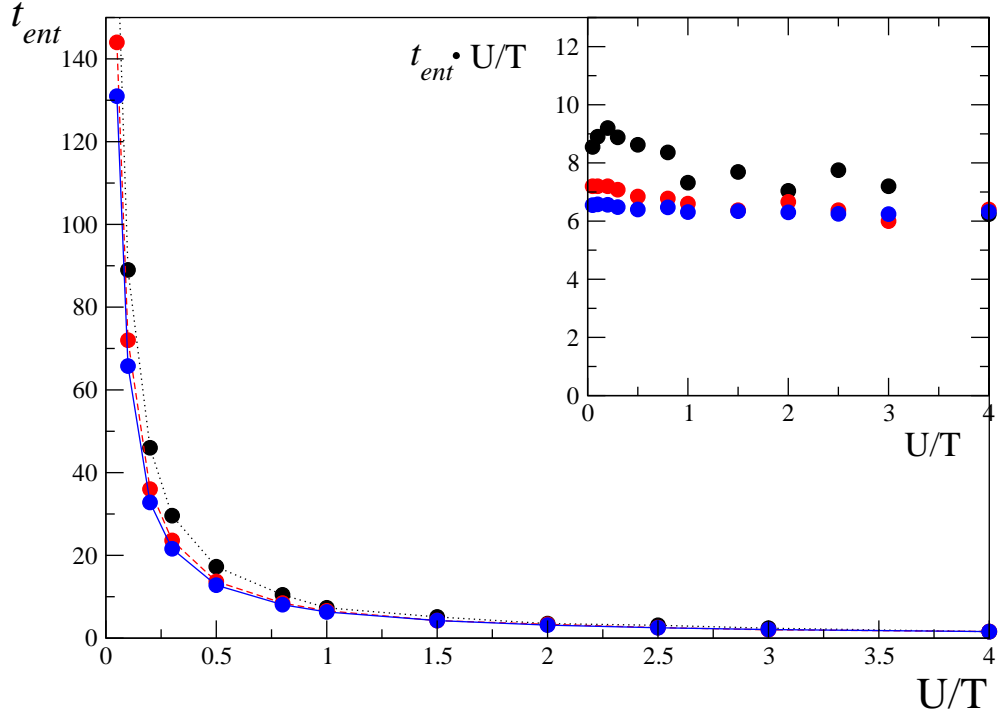}\vspace*{-0.5em}
\end{center}
\caption{\label{tent_tri}
Entanglement period $t_{ent}$ versus interaction for $n=4$ (black), $n=10$ (red) and $n=30$ (blue)
interacting bosons in the pool. 
The product $t_{ent} U/T$ versus $U/T$ is shown in the inset.
}
\end{figure}

Figure 4 shows the dependence of the entanglement period
$t_{ent}$ on the bosonic interaction; it is rather different from
the bipartite case (see Fig. 2({\it b})). Importantly, for small $U/T$ the
entanglement period can reach values orders of magnitude larger than
in the bipartite case, so that the system remains entangled over
extended periods of time. We also note that for sufficiently large
$n$ the dependence of $t_{ent}$ on the interaction scales approximately
as $1/U$, so that $t_{ent}U/T\approx\mathrm{const}$ (see the
inset in Fig. 4). Note that because of the units of $t_{ent}$, the product $t_{ent}U/T$ is $T$-independent.

Next we determine the bipartite entanglement in the system when one
constituent of the tripartite system is traced out. Entanglement measures
for mixed bipartite systems are well defined. In particular, we use the
entanglement of formation and the negativity to quantify the entanglement in
the bipartite system.


\subsection{Bipartite entanglement between the AQDs}

To start with, we trace out the bosonic pool  and consider a mixed two
qubit state described by the reduced density matrix \bea
\rho_1=\left(\begin{array}{cccc} |c_1|^2 & 0 & 0 & 0 \cr 0 & |c_2|^2 &
c_2c_3^* & 0 \cr 0 & c_3c_2^* & |c_3|^2 & 0 \cr 0 & 0 & 0 & |c_4|^2
\end{array} \right)
\label{rho_mix} \ea written in a standard two-qubit basis. Entanglement
of formation $E_f$ can be defined in this case as
\cite{Wootters98}
\beq E_f(\rho_1)=f\left(\frac{1+\sqrt{1-C(\rho_1)^2}}{2}\right)
\label{ent_mix} \eq
with $f(x)=-x\log_2x-(1-x)\log_2(1-x)$, and the concurrence
\beq C(\rho_1)=max\{0,\lambda_1-\lambda_2-\lambda_3-\lambda_4\}.
\label{conc_mix} \eq
Here, $\lambda_i$ are the square roots of the eigenvalues of
$\rho_1\tilde \rho_1$ in descending order, where $\tilde
\rho_1=(\sigma_y\otimes \sigma_y)\rho_1^*(\sigma_y\otimes \sigma_y)$, and
$\rho_1^*$ is a complex conjugate of $\rho_1$. After some algebra we obtain
$\lambda_1=0, \lambda_{2,3}=|c_1||c_4|$ and $\lambda_4=2|c_2||c_3|$,
which have still to be ordered. This allows us to calculate the concurrence
and the corresponding entanglement of formation according
to~\eqref{ent_mix}.

An alternative measure of entanglement is the negativity
\cite{Werner02}
\beq \mathcal{N}(\rho_1)=\frac{||\rho_1^{\Gamma}||_1-1}{2}\,.
\label{negativity} \eq
Here, $||\rho_1^{\Gamma}||_1$ is the trace norm of the
partial transpose $\rho_1^{\Gamma}$ of the bipartite mixed state.
Essentially $\mathcal{N}(\rho_1)$ is just the absolute value of the sum of
negative eigenvalues of $\rho_1^{\Gamma}$. After some algebra we
find that only the eigenvalue
$\lambda=(|c_1|^2+|c_4|^2-\sqrt{(|c_1|^2-|c_4|^2)+4|c_2|^2|c_3|^2})/2$
of the partial transpose of the density matrix \eqref{rho_mix}
can be negative.

\begin{figure}[!bt]
\begin{center}
\includegraphics[width=0.9\textwidth]{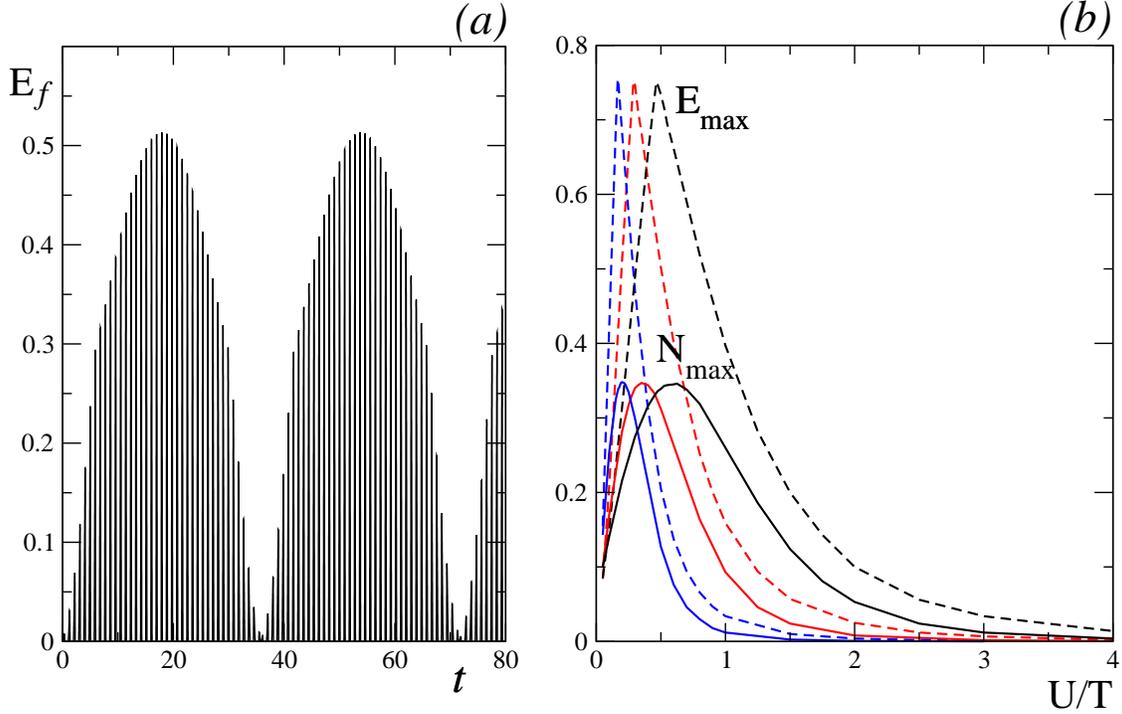}\vspace*{-0.5em}
\end{center}
\caption{
({\it a}) Time dependence of the entanglement of formation $E_f$ for $U/T=0.2$ and $n=10$. ({\it b})
The maximum values of the entanglement of formation $E_{max}$ (dashed curves) and the negativity $N_{max}$ (solid curves)
versus the interaction $U/T$ for
$n=4$ (black), $n=10$ (red) and $n=30$ (blue) bosons in the pool. 
}
\end{figure}

The periodic time dependence of the entanglement of
formation is presented in Fig. 5. We note that the entanglement of formation and
negativity have the same time dependence, however, they are different
in amplitude, which is not surprising since the entanglement of formation and negativity are different entanglement measures \cite{Verstraete,Wei}. We therefore analyze
the dependence of the maximal entanglement of formation $E_{max}$ and
maximal negativity $N_{max}$ on the interaction energy $U/T$. The results
for different boson numbers $n$ are shown in Fig. 5({\it b}).

An unexpected result is that the curves $E_{max}(U)$ and $N_{max}(U)$
are nonmonotonic. Each curve possesses a maximum at a characteristic value of
$U=U_c$. The value of $U_c$ depends on the number of bosons $n$,
more precisely, $U_c$ is increasing with decreasing $n$; on the other hand,
the $E_{max}(U_c)$ and $N_{max}(U_c)$ are constant. The nonmonotonic
behavior of the curves is probably related to the fact that for
$U/T\ll 1$ the main entanglement capacity resides
in the bosonic reservoir, which is traced out in the case at hand.
Indeed, when just one ADQ is traced out instead, we observe a
monotonic dependence (see next section).


\subsection{Bipartite entanglement between a single AQD and the pool}

Finally, we trace out one of the qubits and discuss entanglement of the
hybrid qubit-qutrit system described by the reduced density matrix
\bea \rho_2=\left(\begin{array}{cccc} |c_1|^2 & c_1c_2^* & 0 & 0 \cr c_1^*
c_2 & |c_2|^2 & 0 & 0 \cr 0 & 0 & |c_3|^2 &c_3 c_4^* \cr 0 & 0 & c_3^* c_4 &
|c_4|^2
\end{array} \right).
\label{rho_mix2}
\ea
In order to determine the negativity we again need to find negative
eigenvalues of the partial transpose of the density matrix
\eqref{rho_mix2}. It turns out  that only two eigenvalues 
can be negative, namely
$\lambda_1=(|c_2|^2-\sqrt{|c_2|^4+4|c_3|^2|c_4|^2})/2$ and
$\lambda_2=(|c_3|^2-\sqrt{|c_3|^4+4|c_1|^2|c_2|^2})/2$. We can now calculate
the negativity as an absolute sum of negative eigenvalues of
$||\rho_2^{\Gamma}||_1$ (see previous section).

The entanglement of formation can be also calculated  by doing a convex
expansion of the density matrix $\rho_2$ \cite{Wootters01}: \beq
\rho_2=\alpha \rho_{\alpha}+\beta\rho_{\beta}=\alpha |\Psi_{\alpha}\rangle
\langle \Psi_{\alpha}|+\beta|\Psi_{\beta}\rangle \langle \Psi_{\beta}|,
\label{decomp} \eq whereby the pure states $|\Psi_{\alpha}\rangle$ and
$|\Psi_{\beta}\rangle$ are \bea
|\Psi_{\alpha}\rangle= \frac{1}{\sqrt{\alpha}}\left(c_1|n+1,0\rangle+c_2|n,1\rangle \right), \\
|\Psi_{\beta}\rangle=
\frac{1}{\sqrt{\beta}}\left(c_3|n,0\rangle+c_4|n-1,1\rangle \right). \ea
Here, $\alpha=|c_1|^2+|c_2|^2$ and $\beta=|c_3|^2+|c_4|^2$, so that
$\alpha+\beta=1$. One should note that the decomposition \eqref{decomp}
is unique because of the prohibition against superpositions of states
with different particle numbers. The resultant entanglement of
formation is \cite{Wootters01}
\beq E=\alpha E_{\alpha}+\beta E_{\beta}, \eq
where $E_{{\alpha}(\beta)}$ is the entanglement of formation of the state $
|\Psi_{\alpha}\rangle$ ($|\Psi_{\beta}\rangle$) calculated with
corresponding concurrencies $C_{\alpha}=2|c_1c_2|/\alpha$ and
$C_{\beta}=2|c_3c_4|/\beta$.

Figure 6 shows the entanglement of formation $E_f$ versus
time for a fixed interaction strength; the dependence of the negativity is
similar up to a scaling factor. In Fig. 6({\it b}) the maximum entanglement
of formation $E_{max}$ and maximum negativity $N_{max}$ are
plotted versus interaction $U/T$ for different bosonic occupations. The
behavior of these quantities is monotonic and in line with the
$U/T$-dependence of the maximum concurrence shown in Fig. 3({\it b}):
Smaller boson numbers $n$ in the pool favor larger values of
entanglement. 

\begin{figure}[!bt]
\begin{center}
\includegraphics[width=0.9\textwidth]{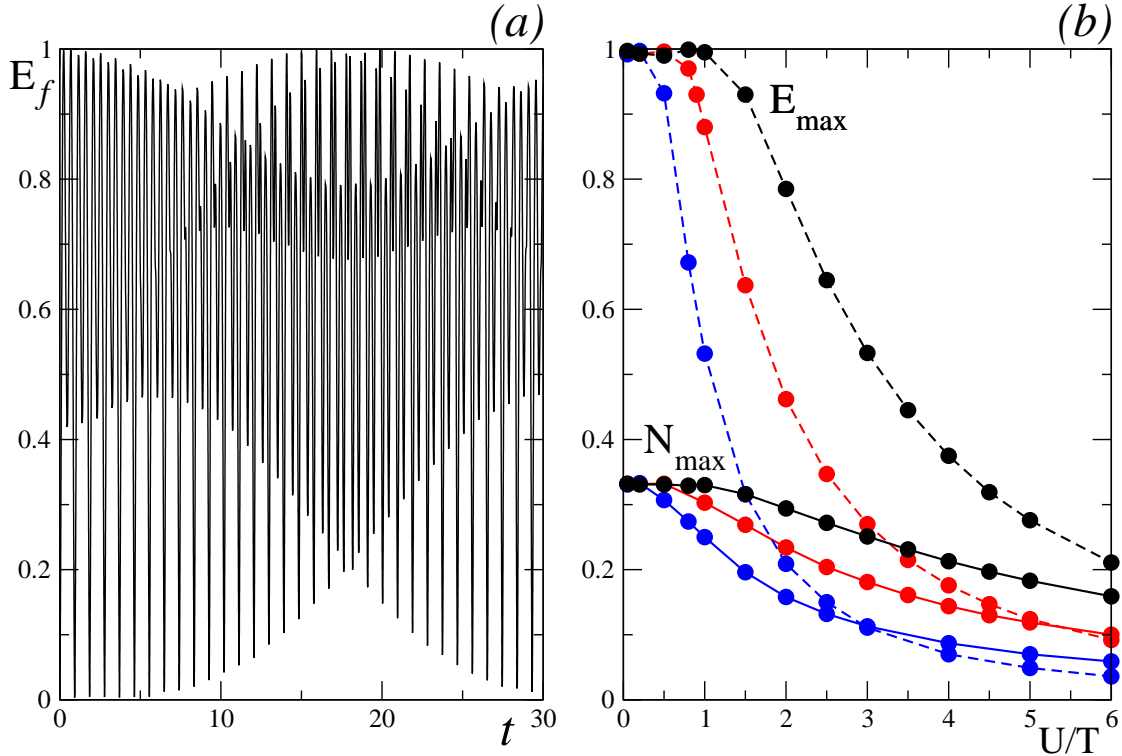}\vspace*{-0.5em}
\end{center}
\caption{
({\it a}) Time dependence of the entanglement of formation $E_f$ for $U/T=0.2$ and $n=10$. ({\it b})
The maximum values of the entanglement of formation $E_{max}$ (dashed curves) and the negativity $N_{max}$ (solid curves)
versus interaction $U/T$ for $n=4$ (black), $n=10$ (red) and $n=30$ (blue)
bosons in the pool. 
}
\end{figure}


\vspace{20pt}

\section{Conclusions and discussions}

We have analyzed qubit-qubit, qubit-qutrit and
qubit-qutrit-qubit systems which can be realized in a cold atomic
set-up. Qutrits have bigger entanglement capacity and offer potentially
more advantages for quantum computation than qubits. However, they are
difficult to handle, as they involve more degrees of freedom.
Qubit-qutrit entanglement has been recently observed for the
first time in a photonic system \cite{Nathan}. Cold atomic
qutrits, which emerge due to the particle conservation, are
characterized by the number of bosons and their interactions; both
parameters can be easily tuned experimentally.

We have demonstrated that both parameters play a crucial role in
generating the maximal possible entanglement between the subsystems as
well as for the longest entanglement period. For example, smaller number of
particles and smaller interactions in general favor the generation of
 entanglement. Note that increasing $n$ ($n\gg 1$) leads to the disappearance of entanglement as expected \cite{EPL}. 
Systems with large interactions
($U/T\gg 1$) do not exhibit entanglement. Interestingly, the
entanglement period of the tripartite system is orders of magnitude larger
than that of the bipartite system. We have also found a simple relation between entanglement period
and interaction for large number of bosons ($n \gg 1$).

Note that in this work we dealt with so-called mode
entanglement \cite{Zanardi}, i.e., entanglement of modes in the Fock
representation. It was debated whether or not mode entanglement can be
effectively used for quantum computatiom because  superselection
rules prohibit coherent superpositions of different particle
number, which introduces severe limitations. However, it has been shown recently that there are ways
to overcome this drawback, for example, mode entanglement can be
used for dense coding \cite{Vedral}.

Our results may also shed light
on the role played by interactions in the problem of coupled
harmonic oscillators \cite{Osc}. One can assume from our findings that
entanglement between oscillators strongly depends on the ratio of
interaction versus coupling.  Another interesting extension  our
results could be the study of entanglement between atomic
Bose-Einstein condensates and immersed impurities \cite{Koehl,Johnson}.

\vspace{1cm}

{\bf Acknowledgments} We are grateful to M. Eschrig, D. Jaksch,
N. Langford, G.J. Milburn,  and W. K. Wootters  for fruitful and valuable comments and discussions.
W.B. acknowledges financial support from the German Research Foundation (DFG) through SFB 767 and SP 1285.

\end{document}